\documentstyle[psfig,11pt,a4]{article}

\def \square {\hbox{$\sqcup\!\!\!\!\sqcap$}} 

\newcommand{\be}{\begin{equation}}
\newcommand{\ee}{\end{equation}} 
\newcommand{\bea}{\begin{eqnarray}}
\newcommand{\eea}{\end{eqnarray}}

\begin{document}

\begin{titlepage}

\begin{flushright} 
{\tt 	FTUV/98-95\\
	IFIC/98-96\\
	SU-ITP/98-66\\
	hep-th/9811246}
 \end{flushright}

\bigskip

\begin{center}

{\bf{\LARGE Can conformal transformations \\
change the fate of  2D black holes?
\footnote{Work partially supported by the 
{\it Comisi\'on Interministerial de Ciencia y Tecnolog\'{\i}a}\/ 
and {\it DGICYT}.}}
}

\bigskip 
\bigskip\bigskip
 J. Cruz$^a$ \footnote{\sc cruz@lie.uv.es}, A. 
 Fabbri$^b$ \footnote{Supported by an INFN fellowship. e-mail: {\sc
afabbri1@leland.stanford.edu}} and
 J. Navarro-Salas$^a$ \footnote{\sc jnavarro@lie.uv.es},

\end{center}

\bigskip%

\footnotesize
	
 a) Departamento de F\'{\i}sica Te\'orica and 
	IFIC, Centro Mixto Universidad de Valencia-CSIC.
	Facultad de F\'{\i}sica, Universidad de Valencia,	
        Burjassot-46100, Valencia, Spain. 
 \newline                 
b) Department of Physics, Stanford University, Stanford, CA, 94305-4060, USA.
\normalsize 

\bigskip

\bigskip

\begin{center}
{\bf Abstract}
\end{center}
 
 By using a classical Liouville-type model of two
 dimensional dilaton gravity we show that the one-loop theory
 implies that the fate of a black hole depends on the conformal frame.
 There is one frame for which the evaporation process never stops and
 another one
 leading to a complete disappearance of the black hole. This can be seen
 as a consequence of the fact that thermodynamic variables are not conformally
 invariant.  In the second case the evaporation always produces the same static
 and regular end-point geometry, irrespective of the initial state.



\end{titlepage}

\newpage

 The understanding of the dynamical evolution of black holes is an important ingredient
 in the formulation of a consistent theory of quantum gravity.
 Since the work of Callan-Giddings-Harvey-Strominger \cite{CGHS}, the study of two-dimensional models
 for black hole formation and evaporation has increased a lot and it has been very useful
 to analyze quantum aspects of  black hole physics. In particular, the existence 
 of exactly solvable one-loop models \cite{RST,BPP} has allowed to study back reaction
 effects in  an analytical setting.
 One of the central properties of the 
 CGHS model is that the Hawking temperature is independent of the mass
 and many aspects of the quantum evolution of the black holes are 
 indeed associated
 with this fact.
 Therefore it is interesting to consider other models giving rise to black hole
 solutions with a more realistic Hawking temperature. In Ref \cite{Cruz} it was analyzed
 a model with a classical gravitational action conformally related to the
  Polyakov-Liouville action
 \be
S={1\over2\pi}\int d^2x\sqrt{-g}\left[R\phi+4\lambda^2e^{\beta\phi}-{1\over2}
\sum_{i=1}^N\left(\nabla f_i\right)^2\right]\>.\label{expcl}
\ee
This model, closely related to the one introduced by Mann \cite{Mann}, has
solutions which resemble  the string 2D black holes \cite{CGHS}, but the Hawking temperature
depends on the mass.  Moreover it also permits to
 construct an associated solvable semiclassical
model and it was also pointed out
 in \cite{Cruz} that the one-loop solution predicts that the Hawking 
 evaporation never stops.
 
 The black hole solutions of the model (\ref{expcl}) are, in an appropriate Kruskal-type gauge, of the form
 \be
 ds^2=\frac{-dx^+dx^-}{{\lambda^2\beta\over C}+Cx^+x^-}\label{i}
 \>,
 \ee
 \be
 e^{\beta\phi}=\frac{1}{{\lambda^2\beta\over C}+Cx^+x^-}\label{ii}
 \>,
 \ee
  where the constant $|C|$ is proportional to the black hole mass. Although
  we cannot
 recover Minkowski spacetime for any value of $C$,
there is a simple way to get a flat geometry.
 By performing a conformal rescaling
 of the metric
 \be
g_{\mu\nu}\rightarrow\frac{ g_{\mu\nu}}{J(\phi)}
 \>,\label{weyli}\ee
 where
 \be
  J(\phi)={1\over\beta} \left(e^{\beta\phi}-1\right)
  \>,\label{weylii}\ee 
  the new geometry, in conformal gauge $ds^2=-e^{2\rho}dx^+dx^-$, is
\be
ds^2=\frac{-dx^+dx^-}{{1\over\beta}-{\lambda^2\over C}-{C\over\beta}x^+x^-}
\>,\label{expressol}
\ee
and it is clear that for $C=\lambda^2\beta$ we obtain a Minkowskian 
ground state, which was not present in the original model.
\footnote{This is a particular case of a general procedure to construct a
theory with a flat ground state starting from an arbitrary 2D dilaton gravity
theory \cite{Katanaev}.}
Moreover expanding the solutions (\ref{expressol}) around $C=\lambda^2\beta$ we have,
to leading order
\be
ds^2=\frac{-dx^+dx^-}{\beta^{-2}\lambda^{-2}(C-\beta\lambda^2)-\lambda^2x^+x^-}
\>,
\ee
which impies that, for $C\sim\beta\lambda^2$, the solutions are 
similar to the CGHS black hole solutions with 
$M={C\over\beta^2\lambda}-{\lambda\over\beta}$.
The above discussion
serves to 
motivate the analysis of the conformally rescaled model through the
transformation (\ref{weyli}-\ref{weylii}). 
Due to the existence of a classical Minkowski ground state one would
expect that the black holes (\ref{expressol}) may now decay completely
(as, for instance, in \cite{RST} and \cite{BPP}).
It has already been stressed that the thermodynamical variables can depend upon the conformal
frame \cite{Chan} and in this paper we shall explicitly show that 
the evaporation process can be indeed
very different.

In terms of the rescaled metric the action (\ref{expcl})
transforms into
\bea
S&=&{1\over2\pi}\int d^2x\sqrt{-g} \left[R\phi+{\beta e^{\beta\phi}\over 
e^{\beta\phi}-1}(\nabla\phi)^2\right.\nonumber\\&&\left.
+{4\lambda^2\over\beta}e^{\beta\phi}\left(e^{\beta\phi}-1\right)-{1\over2}
\sum_{\i=1}^N(\nabla f_i)^2\right]
\>.\label{expres}
\eea
It is also interesting to note that for $\beta=0$ we recover the CGHS model
after the trivial redefinition $\phi\rightarrow e^{-2\phi}$.
Let us now consider the formation of a black hole by collapse of 
an infalling shock wave at $x^+=x^+_0$.
For $x^+<x^+_0$ the metric is given by
\be
ds^2=\frac{dx^+dx^-}{\lambda^2x^+x^-}
\>,\ee
with Minkowskian coordinates ${\sigma^{\pm}}$ defined by
$ \lambda  x^{\pm}=\pm e^{\pm\lambda\sigma^{\pm}}$.
After the incoming shock-wave, $x^+>x^+_0$, the metric is described as
\be
ds^2=\frac{-dx^+dx^-}{{1\over\beta}-{\lambda^2\over C}-{C\over\beta}(x^++\Delta^+)(x^-+\Delta^-)}
\>,\ee
with 
\be
\Delta^+=x^+_0\left({\lambda^2\beta\over C}-1\right)
\>,\ee
\be
\Delta^-={1\over x_0^+}\left({1\over\lambda^2\beta}-{1\over C}\right)
\>,\ee
and the new asymptotically flat coordinates $ \tilde \sigma^{\pm}$ are defined as 
$\sqrt{{C\over\beta}}\left(x^{\pm}+\Delta^{\pm}\right)=\pm e^{\pm
\sqrt{{C\over\beta}}\tilde \sigma^{\pm}}$.   A simple calculation leads to the following
expression for the stress tensor of the shock-wave
\be
T_{++}^f={C\Delta^-\over\beta}\delta(x^+-x^+_0)\>,
\ee
and, in terms of the asymptotically flat coordinates $\sigma^{\pm}$, we have
\be
T^f_{\sigma^+\sigma^+}=\left({C\over\lambda\beta^2}-{\lambda
\over\beta}\right)\delta(\sigma^+-\sigma^+_0)
\>.\ee
This means that the energy of the wave is $M={C\over\lambda\beta^2}-{\lambda
\over\beta}$, which turns out to be, by energy conservation, equal to the ADM 
mass of the black hole solution (\ref{expressol}).  This expression is in accordance
with the mass formula for generic 2D models \cite{Mann2, Gegenberg}, with
an 
appropriate normalization of the Killing vector $k^{\mu}$ 
at spatial infinity 
 ($k^2=-{C\over\lambda^2\beta}$), and differs
in the constant shift $-{\lambda\over\beta}$ from the mass formula for the
solutions (\ref{i}),(\ref{ii}).
 In fact, using the arguments of \cite{Cadoni} one can show that the mass is
 conformally invariant up to a constant shift (related to the choice of the
ground state \footnote{Such a constant shift was not considered in \cite{Cadoni}
because it was supposed there that the ground state of the rescaled theory is
always obtained by conformally transforming the vacuum of the original one.
This is not what happens here.}).
 We have to note that a different normalization of the Killing vector gives 
rise
to a different
expression for the mass which is not compatible with energy conservation.

The flux of radiation measured by inertial observers at future null infinity
$\tilde\sigma^+\rightarrow\infty$ can be worked out as follows
\be
<T^f_{\tilde\sigma^-\tilde\sigma^-}>=-{N\over24}\left\{\sigma^-,\tilde\sigma^-\right\}
\>,\ee
where $\left\{\sigma^-,\tilde\sigma^-\right\}$ is the Schwartzian derivative.
 One obtains
\be
<T^f_{\tilde\sigma^-\tilde\sigma^-}>=
{NC\over48\beta}\left[1-\left(1+\Delta^-\sqrt{{C\over\beta}}
e^{\sqrt{{C\over\beta}}\tilde\sigma^-}\right)^{-2}\right]\label{H}
\>.\ee
At late times $(\tilde\sigma^-\rightarrow \infty)$, the flux of radiation 
approaches a constant
thermal value with an associated Hawking temperature given, in terms of the mass, by
\be
T_H={1\over2\pi}\left(\lambda+\beta M\right)\>.\label{TH}
\ee
We  must stress that in obtaining this expression we have
 used the normalization of the Killing vector already used
  to compute the mass
and that for small masses $M<<{\lambda\over |\beta |}$ one recovers 
the constant temperature of the CGHS black hole.

We have also to point out that due to the shift $\Delta^+$ relating the
asymptotically flat coordinates $\sigma^+$ and $\tilde\sigma^+$ before and
after the collapse
$\left(e^{\lambda\sigma^+}+\lambda
\Delta^+=\right.$ 
\newline
$\left.\lambda\sqrt{{\beta\over
C}}e^{\sqrt{{C\over\beta}}\tilde\sigma^+}\right)$
there exists also an additional (semiclassical) incoming flux
$<T^f_{\tilde\sigma^+
\tilde\sigma^+}>$. This flux is positive and vanishes at
$\tilde\sigma^+\rightarrow\infty$ and one therefore could 
expect that it does not affect
the evaporation process.
Despite the presence of such a non-vanishing $<T^f_{\tilde \sigma^+
\tilde \sigma^+}>$ we will be able, later, to construct
evaporating solutions where this flux does not appear.

Let us now consider the semiclassical one-loop theory in both models 
(\ref{expcl}) and (\ref{expres}).
 One can construct a solvable semiclassical theory,
 maintaining the classical free field equation
$\partial_+\partial_-(2\rho-\beta\phi)=0$, by adding a particular local counterterm
to the standard non-local Polyakov effective action 
\bea
S&=&{1\over2\pi}\int d^2x\sqrt{-g}\left[R\phi+4\lambda^2e^{\beta\phi}-{1\over2}
\sum_{i=1}^N\left(\nabla f_i\right)^2\right]\nonumber\\
&&-{N\over96\pi}\int d^2x\sqrt{-g}\left[R\square^{-1}R+
\beta(2R\phi-\beta(\nabla\phi)^2)\right]
\>.\label{expsemi}
\eea
We have introduced a slightly different counterterm to that of \cite{Cruz}
because now we can also preserve the remaining 
unconstrained classical equation of motion (see (\ref{xix})).
For the theory defined by the classical action  (\ref{expres}) 
a solvable one-loop theory
can be obtained  by transforming  (\ref{expsemi}) with the rescaling
(\ref{weyli}-\ref{weylii}). The new semiclassical theory recovers
 the BPP model \cite{BPP} in the
limit $\beta=0$.
In Kruskal coordinates $(2\rho=\beta\phi)$ the  
equations of motion derived from (\ref{expsemi}) leads to the Liouville equation for the field
$2\rho$
\be
\partial_+\partial_-2\rho=-\lambda^2\beta e^{4\rho}\>,\label{xix}
\ee
 and the constraint equations 
\be
e^{2\rho}\partial_{\pm}^2e^{-2\rho}=\beta\left(T_{\pm\pm}^f-{N\over12}t_{\pm}\right)\>,
\ee
where $t_{\pm}(x^{\pm})$ are the boundary contributions coming from the non-local
 Polyakov term. The simplest solution can be obtained when
$T^f_{\pm\pm}=0=t_{\pm}$
\be
 ds^2={-dx^+dx^-\over{\lambda^2\beta\over C}+Cx^+x^-}\>,\label{thermal}
\ee
and, for $C<0$ and $\lambda^2\beta<0$, the metric (\ref{thermal}) represents
 a black hole in thermal equilibrium.
  As it was illustrated in \cite{Cruz} for the CGHS case we can
   study the evaporation process of the black hole
   by considering the dynamical evolution 
of the solution (\ref{thermal}) when the incoming 
flux is turned off at $x^+=x^+_0$. This can be exactly achieved in the present
theory by assuming the following boundary conditions
\be
t_{x^+}={1\over4(x^++\Delta^+)^2}\Theta(x^+-x^+_0)\>,\label{bfi}
\ee
\be
t_{x^-}=0
\>,
\ee
where
\be
\Delta^+=-x^+_0+{1\over\lambda}\left({C x_0^+\lambda n^{1\over2}
\over C+\lambda\beta^2 M}\right)^{{2\over n+1}}\>,
\ee
and 
\be
n^2=1-{N\beta\over12}\>.\label{ndef}
\ee
The evaporating solution for $x^+>x^+_0$ is
\be
ds^2={-dx^+dx^-\over {\beta\lambda^2 (\lambda x^++\lambda\Delta^+)^{{1-n\over2}}
\over n^{1\over2}(C+\lambda\beta^2 M)}+{(C+\lambda\beta^2 M) 
(\lambda x^++\lambda \Delta^+)^{{n+1\over2}}(\lambda x^-+\lambda\Delta^-)\over
\lambda^2 n^{1\over2}
}}\>,\label{tbr}
\ee
where
\be
\Delta ^-={\lambda^2\beta\over C^2x_0^+}\left[1-{\left(
\lambda x^+_0\right)^{{1-n\over1+n}}\over n^{{n\over1+n}}}\left(
{C\over C+\lambda\beta^2M}\right)^{{2\over1+n}}\right]\>,
\ee
 and there is additionally a shock wave along $x^+=x_0^+$
 \be
 T^f_{++}=\left[{\lambda^{{1-n\over1+n}}(n+1)\over2\beta}\left(
 {C+\lambda\beta^2 M\over Cx^+_0n^{1\over2}}\right)^{{2n\over1+n}}-
 {1\over\beta x^+_0}\right]\delta (x^+-x^+_0)\>.
 \ee
 This shock wave can be eliminated with an appropriate
  choice of the parameter
 $M$ ($M$ turns out to be the energy of the shock wave in the classical limit).
 The main property of the evaporating solution
  is that the apparent horizon ($\partial_+\phi=0$)
\be
\lambda x^-=-\lambda\Delta^-+{\lambda^4\beta (n-1)\over 
(C+\lambda\beta^2 M)^2(n+1)
(\lambda x^++
\lambda\Delta^+)^n}\>,\label{apphor}
\ee
and the singularity curve 
\be
\lambda^4\beta+(C+\lambda\beta^2 M)^2(\lambda x^++\lambda\Delta^+)^n(\lambda
x^-+\lambda\Delta^-)=0
\>,
\ee
 never meet each other, implying 
 that the Hawking evaporation never ceases.
 
The point now is to see what happens in the evaporation process
 of the conformally rescaled model with $\lambda^2\beta<C<0$, which is the range of
 variation of $C$ valid for both models.
 The rescaled dynamical solution, after switching 
 off the incoming flux at $x^+>x^+_0$ is
\bea
ds^2&=&-\left[{1\over\beta}-{\lambda^2(\lambda x^++
\lambda\Delta^+)^{{1-n\over2}}
\over n^{1\over2}(C+\lambda\beta^2 M)}\right.\nonumber\\&&\left.
-{C+\lambda\beta^2 M\over n^{1\over2}
\lambda^2\beta}
(\lambda x^++\lambda \Delta^+)^{{n+1\over2}}(\lambda x^-+\lambda\Delta^-)
\right]^{-1}dx^+dx^-
\>.\label{expressolsemi}
\eea
The curvature singularity of (\ref{expressolsemi}) 
\bea
&n^{1\over2}\lambda^2(C+\lambda\beta^2 M)(\lambda x^++\lambda\Delta^+)^{{n-1\over2}}-
\lambda^4\beta&\nonumber\\&-(C+\lambda\beta^2 M)^2(\lambda
x^++\lambda\Delta^+)^n(\lambda
x^-+\lambda\Delta^-)=0&
\>,
\eea
is hidden behind the apparent horizon (\ref{apphor}) immediately after 
$x^+=x^+_0$, but they intersect each other at the point
\be
x^+_{int}=-\Delta^++{1\over\lambda}\left({2n^{1\over2}\lambda^2\beta\over 
(n+1)(C+\lambda \beta^2 M)}\right)^{{2\over n-1}}\>,
\ee
\be
x^-_{int}= -\Delta^-+{\lambda^3\beta(n-1)\over(C+\lambda\beta^2 M)^2(n+1)}
\left({2n^{1\over2}\lambda^2\beta\over(C+\lambda\beta^2
M)(n+1)}\right)^{{2n\over1-n}}\>.
\ee
Remarkably, the evaporating solution can be matched 
at the end-point $x^-=x^-_{int}$ with a static stable and regular 
solution.

 To show this let us first analyze the static
radiationless solutions of the model.
With the following boundary conditions
\be
t_{x^\pm}={1\over4(x^{\pm})^2}\>,\label{bf}
\ee
one can check that the  solutions 
\be
ds^2=\frac{-dx^+dx^-}{{1\over\beta}-{\lambda^2\over n C}
(-\lambda^2x^+x^-)^{{1-n\over2}}+{C\over n\lambda^2\beta}
(-\lambda^2 x^+x^-)^{{n+1\over2}}}
\>,\label{static}
\ee
where
$C$ is an integration constant and 
$n$ is given by (\ref{ndef}),
 are stable with respect to the asymptotically Rindlerian
 coordinates  $\sigma^{\pm}$ defined by
 \be
 \lambda x^{\pm}=\pm e^{\pm {C\over\lambda\beta} \sigma^{\pm}}.
 \ee
 The requirement of stability with respect to these coordinates
seems natural because   in terms 
of them the solutions (\ref{static}) are independent of the time coordinate
( similar considerations using asymptotically Rindler coordinates
in solvable models of 2D gravity can also be found in \cite{Fabbri1} ). 
For $C<\hat C$, where 
\be
 \hat C=-\lambda^2(-\beta)^{{n+1\over2}}{2^n\over
 (n+1)^{{n+1\over2}}(n-1)^{{n-1\over2}}}\>,
 \ee
the spacetime geometry has a naked singularity, for $C=\hat C$ the solution
is completely regular and for $C>\hat C$ there are null singularities at
$x^+x^-=0$.
Alternatively one can also choose these solutions to be stable with respect to
the asymptotically Minkowskian frame although in these coordinates the
solutions (\ref{static}) are no longer static (the same prescription was
adopted in \cite{Fabbri2}) \footnote{We must point out that 
 in the classical limit both descriptions coincide.}.
In this case  the boundary functions (\ref{bf}) must  be substituted by
\be
t_{x^{\pm}}={2n-n^2+3\over16(x^{\pm})^2}\>,
\ee
and $n$ is now given by
\be
 n=\frac{2-{N\beta\over8}}{2-{N\beta\over24}}\>.
 \ee
Analogously, in order to eliminate the incoming flux in the evaporating
solution (\ref{tbr}) the boundary function (\ref{bfi}) would be substituted
by 
\be
t_{x^{+}}={2n-n^2+3\over16(x^{+}+\Delta^+)^2}\Theta (x^+-x^+_0)\>.
\ee  

As we have already mentioned
the evaporating solution  (\ref{expressolsemi}) can be  just matched with the
static and regular solution with $C=\hat C$
\bea
ds^2&=&-\left[{1\over\beta}-{\lambda^2\over n\hat C}(\lambda
x^++\lambda\Delta^+)^{{1-n\over2}}(-\lambda x^--\lambda\tilde\Delta^-)^{{1-n
\over2}}\right.\nonumber\\&&\left.
+{\hat C\over n\lambda^2\beta}(\lambda 
x^++\lambda\Delta^+)^{{n+1\over2}}(-\lambda
x^--\lambda\tilde\Delta^-)^{{n+1\over2}}\right]^{-1}dx^+dx^-\>,\label{staticii}
\eea
where
\be
\tilde\Delta^-=n\Delta^-+(n-1)x^-_{int}
\>,
\ee
and there is emission of a``thunderpop" along the null line
$x^-=x^-_{int}$
\be
T^f_{--}={1-n\over2n\beta(x^-_{int}+\Delta^-)}\delta(x^--x^-_{int})\>.
\ee

We can obtain the same result analizing the evaporation process of a black 
hole
formed by gravitational collapse. The static radiationless solutions 
(\ref{static}) ($C \le \hat C$) can be matched 
at the shock
wave line $x^+=x^+_0$ with an evaporating solution
\be
ds^2=\frac{-dx^+dx^-}{{1\over\beta}-{\lambda^2((\lambda x^++\lambda \Delta^+)
(-\lambda x^-))^{{1-n\over2}}
\over n(C+\lambda\beta^2 M)}+{C+\lambda\beta^2 M\over n\lambda^2\beta}{
(\lambda x^++\lambda \Delta^+)^{{n+1\over2}}((-\lambda x^-)^n-\lambda\Delta^-)\over 
(-\lambda x^-)^{{n-1\over2}}}}
\>,\label{semires}
\ee
with 
\be
\Delta^+=x_0^+\left[\left({C\over C+\lambda\beta^2 M}\right)^{{2\over1+n}}
-1\right]\>,
\ee
\be
\Delta^-={\lambda^3\beta\over (\lambda x_0^+)^nC^2}\left[1-
\left({C\over C+\lambda\beta^2 M}\right)^{{2\over1+n}}\right]
\>,
\ee
and the energy momentum tensor takes the expression
\be
T^f_{++}={n+1\over2\beta x_0^+}\left[\left({C+\lambda\beta^2 M\over
C}\right)^{{2\over n+1}}-1\right]\delta(x^+-x^+_0)\>.
\ee
The singularity curve of (\ref{semires})
\bea
&n\lambda^2(C+\lambda\beta^2 M)((\lambda x^++\lambda\Delta^+)(-\lambda x^-))
^{{n-1\over2}}-
\lambda^4\beta&\nonumber\\&+(C+\lambda\beta^2 M)^2(\lambda x^+
+\lambda\Delta^+)^n((-\lambda
x^-)^n-\lambda\Delta^-)=0&\>,\label{singsemi}
\eea
is hidden behind the apparent horizon
\be
(-\lambda x^-)^n=\lambda\Delta^-+{\lambda^4\beta (1-n)\over 
(C+\lambda\beta^2 M)^2(1+n)
(\lambda x^++
\lambda\Delta^+)^n}
\>,\label{ahsemi}
\ee
provided  the mass
 $M$ is above a critical mass $M_{cr}$ (vanishing for $C=\hat C$),
  and the black hole shrinks until the
 intersection point
 \bea
x^+_{int}&=&-\Delta^++\lambda^{-{1+n\over n}}(\Delta^-)^{-{1\over n}}
\left[\left({2\lambda^2\beta\over( C+\lambda\beta^2 M)(1+n)}
\right)^{{2n\over n-1}}\right.\nonumber\\&&\left.
-{\lambda^4(1-n)\over (C+\lambda\beta^2 M)^2(1+n)}\right]^{1\over n}
\>,
\eea
\bea
x^-_{int}&=&-{(\lambda\Delta^-)^{{1\over n}}\over\lambda}\left(
{\lambda^2\beta\over(C+\lambda\beta^2
M)(1+n)}\right)^{{2\over n-1}}                                    
\left[\left({2\lambda^2\beta\over( C+\lambda\beta^2 M)(1+n)}
\right)^{{2n\over n-1}}\right.\nonumber\\&&\left.
-{\lambda^4(1-n)\over (C+\lambda\beta^2 M)^2(1+n)}\right]^{-{1\over n}}
\>.
\eea
The evaporating solution can also be matched across $x^-=x^-_{int}$ with the
regular static solution (\ref{staticii}) where now
\be
 \tilde\Delta^-=\Delta^-(-\lambda x^-_{int})^{{1-n}}\>,
\ee
and with emission of a "thunderpop" at $x^-=x^-_{int}$
\be
T^f_{--}={1-n\over2\beta x^-_{int}}\left[\left({C+\lambda\beta^2 M\over\hat
C}\right)^{{2\over1-n}}-1\right]\delta(x^--x^-_{int})\>.
 \ee

We want to remark that the  
different result for the process of black hole evaporation in the initial
Liouville-type model (\ref{expcl}) and in the rescaled one 
(\ref{expres}) can be seen as 
a consequence 
of the different relation between thermodynamic variables in each of these
models.
In  the model (\ref{expcl}) the temperature is proportional to the
 black hole mass 
 and therefore goes to zero when the black hole mass
 becomes very small preventing the complete evaporation.
On the contrary in the model (\ref{expres}) due to the shift in 
the temperature (\ref{TH}) it approaches a constant non-vanishing
value ${\lambda\over2\pi}$ when the black hole mass goes to zero and
consequently the black hole evaporates completely.
In this second model the evaporation always ends with the same remnant
geometry, irrespective of the initial state or the type of evaporation process
(thermal bath removal or gravitational collapse). We find that this end-point
geometry is everywhere regular, as in other models \cite{RST,BPP}
for which the evaporation process can be followed analytically,
suggesting an underlying general behaviour.
Moreover the exact solvability of these models can be used to analyze
other physical aspects in an analytical setting
(critical behaviour, thermality, etc).
We will consider these questions and details of this work in a future
publication.

 \section*{Acknowledgements}
 J. C. acknowledges the Generalitat Valenciana for a FPI fellowship.
 J. C. and J. N-S. want to thank P. Navarro, J. M. Izquierdo and A. Mikovic
  for useful discussions.
 A. F. wishes to thank N. Kaloper for
interesting remarks.


\begin{thebibliography}{99}

\bibitem{CGHS}
	C. G. Callan, S. B. Giddings, J. A. Harvey and A. Strominger,
	{\it Phys. Rev.}\/ D45 (1992) 1005.
\bibitem{RST}
	J. G. Russo, L. Susskind and L. Thorlacius, {\it Phys. Rev.}\/ 
	D46 (1993) 3444; {\it Phys. Rev.}\/ D47 (1993) 533.
\bibitem{BPP}
	S.  Bose, L.  Parker and Y.  Peleg, 
	{\it Phys. Rev.}\/ D52 (1995) 3512 ; {\it Phys. Rev. Lett.} 
	76 (1996) 861.
	
\bibitem{Cruz}
	J.~Cruz and J.~Navarro-Salas, {\it Phys. Lett.} B387 (1996) 51.
\bibitem{Mann}
 R. B. Mann,  {\it Nucl. Phys.}  B418 (1994) 231.
 
 \bibitem{Katanaev}
 M. O. Katanaev, W. Kummer and H. Liebl, {\it Nucl. Phys.} B486 (1997) 353.
 
 \bibitem{Chan}	
	K. C. K. Chan, gr-qc/9701029.
	\newline
	K.  C.  K.  Chan, J. D. E. Creighton and R. B. Mann, {\it 
	Phys. Rev.} D54 (1996) 3892.
	\newline
	H. Liebl, D. V. Vassilevich and S. Alexandrov, 
       {Class. Quant. Grav.} 14 (1997) 889.

\bibitem{Mann2}
R. B. Mann, {\it Phys. Rev.} D47 (1993) 4438.

\bibitem{Gegenberg}
J. Gegenberg, G. Kunstatter and D. Louis-Martinez, {\it Phys. Rev.} D51
(1995) 1781.

\bibitem{Cadoni}
M. Cadoni, {\it Phys. Lett.} B395 (1997) 10.

\bibitem{Fabbri1}
A. Fabbri and J. G. Russo, {\it Phys. Rev.} D53 (1996) 6995.

\bibitem{Fabbri2}
R. Balbinot and A. Fabbri, {\it Class. Quant. Grav.} 14 (1997) 463.

\end{thebibliography}
\end{document}